\documentclass[11pt]{article}
\usepackage{amsmath,amssymb,amsfonts,amsthm,bbm,mathrsfs,tcolorbox,tikz}
\usepackage{natbib}
\usepackage{cancel,sectsty,color,enumitem,array,lscape,hyperref,graphicx,subfig}
\usepackage[singlespacing]{setspace}
\usepackage[justification=centering]{caption}
\hypersetup{pdfborder = {0 0 0},colorlinks=true,linkcolor=blue,urlcolor=blue,citecolor=blue}
\usepackage[normalem]{ulem}

\usepackage[top=1in,bottom=1in,left=1in,right=1in]{geometry}

\newtheoremstyle{exampstyle}
  {} % Space above
  {} % Space below
  {} % Body font
  {} % Indent amount
  {\bfseries\color{blue}} % Theorem head font
  {.} % Punctuation after theorem head
  {.5em} % Space after theorem head
  {} % Theorem head spec (can be left empty, meaning `normal')
\theoremstyle{exampstyle}
\theoremstyle{exampstyle}\newtheorem{thm}{Theorem}
\theoremstyle{exampstyle}\newtheorem*{thm*}{Theorem}
\theoremstyle{exampstyle} 
\theoremstyle{exampstyle}\newtheorem{Assumption}{Assumption}
\theoremstyle{definition}

\allowdisplaybreaks[1]

\allowdisplaybreaks[1]

\definecolor{gray1}{rgb}{0.4,0.4,0.4}
\definecolor{gray2}{rgb}{0.8,0.8,0.8}

\newcommand{\ContextOne}{\mathtt{I}}
\newcommand{\ContextTwo}{\mathtt{II}}
\newcommand{\RandomConsideration}{\mathcal{O}}
\begin{document}

% % % % % % % % % % % % % % Title Page % % % % % % % % % % % % % %
\title{Context-Dependent Heterogeneous Preferences: A Comment on Barseghyan and Molinari (2023)
\bigskip}
\author{Matias D. Cattaneo\thanks{Department of Operations Research and Financial Engineering, Princeton University.}\and
        Xinwei Ma\thanks{Department of Economics, UC San Diego.}\and
        Yusufcan Masatlioglu\thanks{Department of Economics, University of Maryland.}}
\maketitle

\begin{abstract}
    \citet{Barseghyan-Molinari_2023_JBES} give sufficient conditions for semi-nonparametric point identification of parameters of interest in a mixture model of decision-making under risk, allowing for unobserved heterogeneity in utility functions and limited consideration. A key assumption in the model is that the heterogeneity of risk preferences is unobservable but context-independent. In this comment, we build on their insights and present identification results in a setting where the risk preferences are allowed to be context-dependent.
\end{abstract}

\thispagestyle{empty}
\clearpage

\onehalfspacing
\setcounter{page}{1}
\pagestyle{plain}

\clearpage

%%%%%%%%%%%%%%%%%%%%%%%%%%%%%%%%%%%%%%%%%%%%%%%
%%%%%%%%%%%%%%%%%%%%%%%%%%%%%%%%%%%%%%%%%%%%%%%
%% INTRODUCTION
%%%%%%%%%%%%%%%%%%%%%%%%%%%%%%%%%%%%%%%%%%%%%%%
%%%%%%%%%%%%%%%%%%%%%%%%%%%%%%%%%%%%%%%%%%%%%%%

\section{Introduction}\label{section:Introduction}

\citet{Barseghyan-Molinari_2023_JBES} offer identification results of risk preferences based on observed bundle choices of decision makers in insurance markets. (See also \citet*{Barseghyan-Molinari-Thirkettle_2021_AER} for related work and background references.) Their economic model allows for multiple preference types, unobserved heterogeneity within each type, and unobserved heterogeneity in (random) consideration sets at the bundle level. (See \citet*{Cattaneo-Ma-Masatlioglu-Suleymanov_2020_JPE}, \citet*{Barseghyan-Coughlin-Molinari-Teitelbaum_2021_ECMA}, and \citet{Cattaneo-Ma-Masatlioglu-Suleymanov_2023_wp} for background references on random attention and related models.) In particular, \citet{Barseghyan-Molinari_2023_JBES} consider decision-making under uncertainty for bundle choices (e.g., collision and comprehensive auto insurance deductibles), allowing for different utility models via a finite mixture of preference types, where each preference type is parameterized with random coefficients. The mixing probabilities for different types are context-independent; that is, for each decision maker, the same utility function is employed in all contexts. 

\citet{Barseghyan-Molinari_2023_JBES}'s key identification insight is to exploit a single-crossing property of the utility models, together with the assumption that prices enter the utility calculation but variations thereof are independent of the preference type, the random utility parameter, and the consideration set formation. Then, assuming there exists sufficient variation in prices, they are able to ``match'' decision makers of different preference types to marginal price changes in different contexts, and semi-nonparametric point identification of the parameters of interest (i.e., the share of preference types and distributions of the random utility parameters) can be achieved from observed choice bundles only.

A key assumption in their model is that the heterogeneity of risk preferences is unobservable but context-independent (in other words, risk behaviors are consistent across environments). While this assumption is the central tenet of classical behavioral models under risk, a large body of evidence documents robust evidence for context-dependent risk behavior \citep{Camerer_1995_HandbookCh,Camerer_1998_EE}. For example, individuals can be seen as risk-averse for gambles involving significant gains or small losses and risk-seeking for gambles involving small gains or significant losses, also known as the ``fourfold pattern'' of risk preferences \citep{Markowitz_1952_JPE,Tversky-Kahneman_1992_JRU}. Moreover, \citet{MacCrimmon-Wehrung_1986_BookCh,MacCrimmon-Wehrung_1990_MS} documented that the degree of risk-taking of the same individual is influenced by decision environments such as games of chance/gambling, financial investing, business decisions, and personal decisions. Hence, it is more natural to allow for risk preferences to be malleable and domain-specific \citep{Weber-Blais-Betz_2002_JBDM}.

Motivated by the aforementioned theoretical and empirical evidence from behavioral sciences, in this comment, we enhance the model of \citet{Barseghyan-Molinari_2023_JBES} to allow for context-dependent risk behavior by permitting the mixing probability entering the finite mixture of preference types to be context-dependent. To be precise, our model introduces a new type of decision makers who employ different utility functions (and, therefore, different random utility parameters) for risk assessment across contexts. Due to the presence of such decision makers, the mixed derivative of the observed choice probability with respect to price variations in different contexts will not be zero, meaning that we can no longer ``match'' decision makers of different types to price variations in different contexts.

To achieve semi-nonparametric identification, we build on the insight of \citet{Barseghyan-Molinari_2023_JBES}, and observe that context-independent preference may lead to non-smooth responses to price variations. To provide some intuition, consider a decision maker who uses the same utility function for risk assessment in all contexts, and she chose products with high deductibles (i.e., low prices). As price variations across contexts are not perfectly correlated, products in some contexts can be considered ``cheap,'' while in other contexts are more ``expensive.'' The decision maker will only react to price changes in contexts where the costs are already low. In other words, her decision to purchase high-deductible products cannot be simultaneously binding in all contexts. On the other hand, if she employs different utility functions across contexts, then it is possible that all her choices are binding, in which case the choice behavior will react to price variations in all contexts. We thus present an identification strategy based on discontinuity in derivatives. 

The remainder of this comment proceeds as follows. Section \ref{sec: Identification Idea under Full Consideration} reviews the key identification insights from \citet{Barseghyan-Molinari_2023_JBES} under full attention, and then presents identification results for context-dependent preferences leveraging those insights. Section \ref{sec: Context-Dependent Mixing Probabilities under Limited Consideration} extends our identification results for context-dependent preferences to settings with random and limited consideration. Section \ref{sec: Conclusion} concludes.

%%%%%%%%%%%%%%%%%%%%%%%%%%%%%%%%%%%%%%%%%%%%%%%
%%%%%%%%%%%%%%%%%%%%%%%%%%%%%%%%%%%%%%%%%%%%%%%
%% Identification under Full Consideration
%%%%%%%%%%%%%%%%%%%%%%%%%%%%%%%%%%%%%%%%%%%%%%%
%%%%%%%%%%%%%%%%%%%%%%%%%%%%%%%%%%%%%%%%%%%%%%%
\section{Identification under Full Consideration}\label{sec: Identification Idea under Full Consideration}

We employ the notation in \cite{Barseghyan-Molinari_2023_JBES} with minimal modifications, and we also adopt their assumptions throughout this comment with the exception of their Assumption 2.2, which we aim to generalize.

%%%%%%%%%%%%%%%%%%%%%%%%%%%%%%%%%%%%%%%%%%%%%%%
%% Model and Identification Insight
%%%%%%%%%%%%%%%%%%%%%%%%%%%%%%%%%%%%%%%%%%%%%%%
\subsection{Model and Identification Insight}

We assume there are two contexts (i.e., choice problems), indexed by $\ContextOne$ and $\ContextTwo$, and the decision maker has to choose between two (risky) alternatives within each context. We label a decision maker by $i$, and the prices she faces in the two contexts are $\mathsf{x}^\ContextOne_i$ and $\mathsf{x}^\ContextTwo_i$, respectively. As we show below, the identification of the parameters will rely on exogenous variation of prices. The decision maker's utility function can be either $U_{\nu_i}(\cdot)$ or $U_{\omega_i}(\cdot)$, with probability $\alpha$ and $1-\alpha$, respectively. The risk preference parameters, $\nu_i$ and $\omega_i$, are realized from distributions $F$ and $G$, respectively, with supports $[0,\bar{\nu}]$ and $[0,\bar{\omega}]$.

Given the price $\mathsf{x}^\ContextOne_i$ (or $\mathsf{x}^\ContextTwo_i$), and the assumptions imposed by \cite{Barseghyan-Molinari_2023_JBES}, there exists a unique risk preference parameter value such that the decision maker is indifferent between the two options. Formally, we define for type-$\nu$ decision makers:
\begin{alignat*}{2}
 \nu_i&\leq \mathcal{V}^{1,1}_{2,1}(\mathsf{x}^\ContextOne_i)&&\qquad \Leftrightarrow\qquad \text{bundle $\mathcal{I}_{1,1}$ is preferred to $\mathcal{I}_{2,1}$ with utility $U_{\nu_i}(\cdot)$},\\
 \nu_i&\leq\mathcal{V}^{1,1}_{1,2}(\mathsf{x}^\ContextTwo_i)&&\qquad \Leftrightarrow\qquad \text{bundle $\mathcal{I}_{1,1}$ is preferred to $ \mathcal{I}_{1,2}$ with utility $U_{\nu_i}(\cdot)$},
\end{alignat*}
which is possible when the utility function exhibits single crossing property \citep[Assumption 2.4]{Barseghyan-Molinari_2023_JBES}. We recall that $\mathcal{V}^{\ell,q}_{k,r}(\cdot)$ denotes the cutoff level for $\nu_i$
at which the agent is indifferent between bundles $\mathcal{I}_{\ell,q}$ and $\mathcal{I}_{k,r}$. In general, the cutoff value would depend on both prices, $\mathsf{x}^\ContextOne_i$ and $\mathsf{x}^\ContextTwo_i$, but thanks to the ``narrow bracketing'' assumption \citep[Assumption 2.3]{Barseghyan-Molinari_2023_JBES}, $\mathcal{V}^{1,1}_{2,1}(\cdot)$ is only a function of $\mathsf{x}^\ContextOne_i$, and $\mathcal{V}^{1,1}_{1,2}(\cdot)$ is only a function of $\mathsf{x}^\ContextTwo_i$. Similarly, we can also define the cutoff values for type-$\omega$ decision makers:
\begin{alignat*}{2}
 \omega_i&\leq \mathcal{W}^{1,1}_{2,1}(\mathsf{x}^\ContextOne_i)&&\qquad \Leftrightarrow\quad \text{bundle $\mathcal{I}_{1,1}$ is preferred to $\mathcal{I}_{2,1}$ with utility $U_{\omega_i}(\cdot)$},\\
 \omega_i&\leq \mathcal{W}^{1,1}_{1,2}(\mathsf{x}^\ContextTwo_i)&&\qquad \Leftrightarrow\quad \text{bundle $\mathcal{I}_{1,1}$ is preferred to $\mathcal{I}_{1,2}$ with utility $U_{\omega_i}(\cdot)$}.
\end{alignat*}

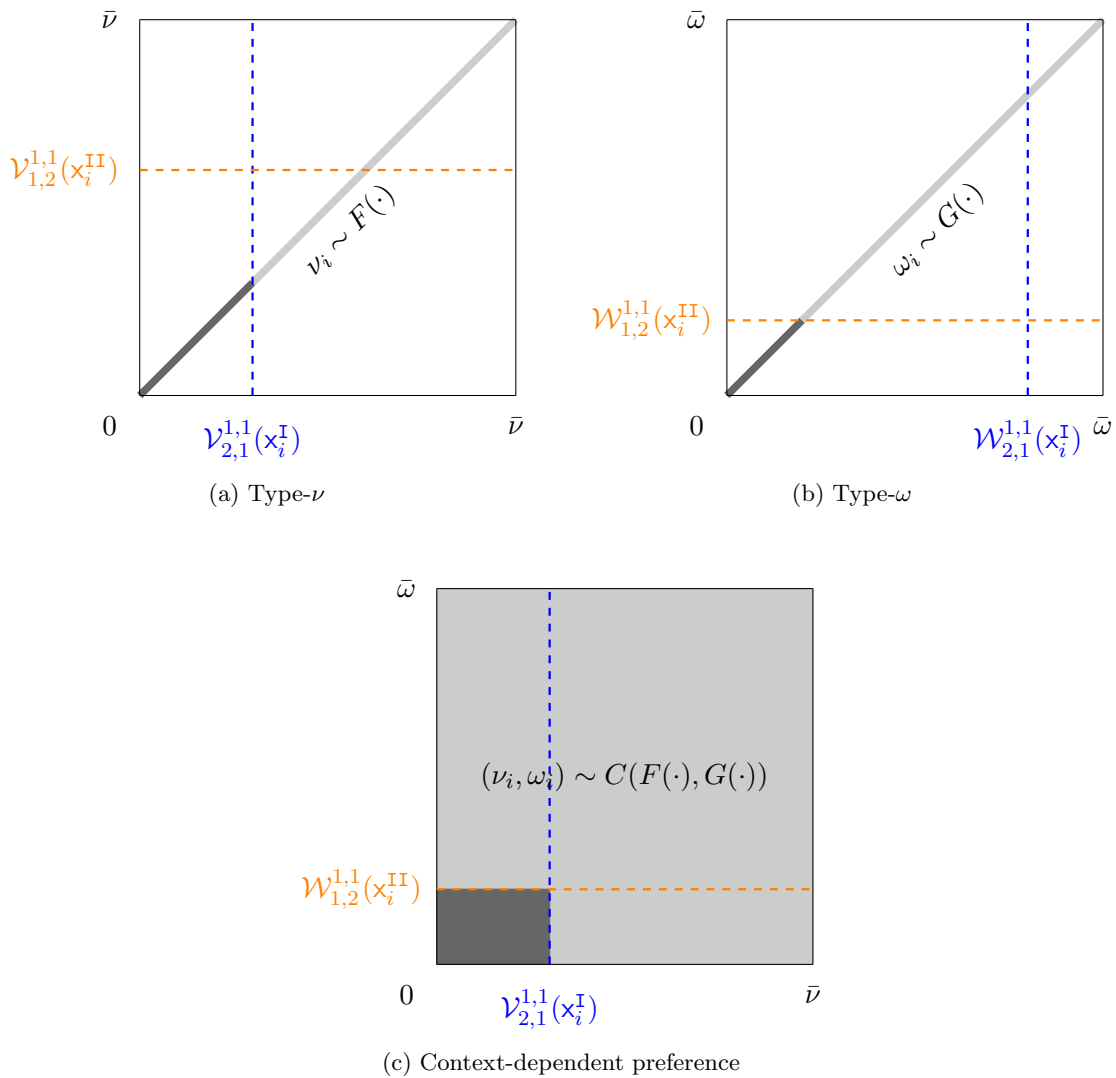
\begin{figure}[!th]
\centering
\subfloat[Type-$\nu$]{
\begin{tikzpicture}

\draw[gray1, line width=1mm] (0, 0)  -- (1.5, 1.5);
\draw[gray2, line width=1mm] (1.5, 1.5)  -- (5, 5);

\draw (0, 0)  -- (0, 5);
\draw (0, 5)  -- (5, 5);
\draw (0, 0)  -- (5, 0);
\draw (5, 0)  -- (5, 5);

\node at (-0.4 , -0.4)  {$0$};
\node at (5 , -0.4)  {$\bar{\nu}$};
\node at (-0.4 , 5)  {$\bar{\nu}$};

\node at (1.5 , -0.6)  {\color{blue}$\mathcal{V}_{2,1}^{1,1}(\mathsf{x}^\ContextOne_i)$};
\node at (-1 , 3)  {\color{orange}$\mathcal{V}_{1,2}^{1,1}(\mathsf{x}^\ContextTwo_i)$};

\draw[dashed, blue, line width=0.3mm] (1.5, 0)  -- (1.5, 5);
\draw[dashed, orange, line width=0.3mm] (0, 3)  -- (5, 3);

\node[rotate=45] at (2.8 , 2.2) {$\nu_i\sim F(\cdot)$};

\end{tikzpicture}
}\quad
\subfloat[Type-$\omega$]{
\begin{tikzpicture}

\draw[gray1, line width=1mm] (0, 0)  -- (1, 1);
\draw[gray2, line width=1mm] (1, 1)  -- (5, 5);

\draw (0, 0)  -- (0, 5);
\draw (0, 5)  -- (5, 5);
\draw (0, 0)  -- (5, 0);
\draw (5, 0)  -- (5, 5);

\node at (-0.4 , -0.4)  {$0$};
\node at (5 , -0.4)  {$\bar{\omega}$};
\node at (-0.4 , 5)  {$\bar{\omega}$};

\node at (4 , -0.6)  {\color{blue}$\mathcal{W}_{2,1}^{1,1}(\mathsf{x}^\ContextOne_i)$};
\node at (-1 , 1)  {\color{orange}$\mathcal{W}_{1,2}^{1,1}(\mathsf{x}^\ContextTwo_i)$};

\draw[dashed, blue, line width=0.3mm] (4, 0)  -- (4, 5);
\draw[dashed, orange, line width=0.3mm] (0, 1)  -- (5, 1);

\node[rotate=45] at (2.8 , 2.2) {$\omega_i\sim G(\cdot)$};

\end{tikzpicture}
}\\ \bigskip
\subfloat[Context-dependent preference]{
\begin{tikzpicture}

\draw [fill=gray2,gray2] (0,0) rectangle (5,5);
\draw [fill=gray1,gray1] (0,0) rectangle (1.5,1);

\draw (0, 0)  -- (0, 5);
\draw (0, 5)  -- (5, 5);
\draw (0, 0)  -- (5, 0);
\draw (5, 0)  -- (5, 5);

\node at (-0.4 , -0.4)  {$0$};
\node at (5 , -0.4)  {$\bar{\nu}$};
\node at (-0.4 , 5)  {$\bar{\omega}$};

\node at (1.5 , -0.6)  {\color{blue}$\mathcal{V}_{2,1}^{1,1}(\mathsf{x}^\ContextOne_i)$};
\node at (-1 , 1)  {\color{orange}$\mathcal{W}_{1,2}^{1,1}(\mathsf{x}^\ContextTwo_i)$};

\draw[dashed, blue, line width=0.3mm] (1.5, 0)  -- (1.5, 5);
\draw[dashed, orange, line width=0.3mm] (0, 1)  -- (5, 1);

\node at (2.5 , 2.5) {$(\nu_i,\omega_i)\sim C(F(\cdot),G(\cdot))$};

\end{tikzpicture}
}
\caption{Choice behavior of different types of decision makers. 
}
\label{fig:fig1}
\begin{flushleft}
Panel (a): Type-$\nu$ decision makers employ $U_{\nu_i}$ for risk assessments in both contexts (context $\ContextOne$ along the horizontal axis and $\ContextTwo$ along the vertical axis). They will choose the bundle $\mathcal{I}_{1,1}$ if the parameter $\nu_i$ is below both $\mathcal{V}_{2,1}^{1,1}(\mathsf{x}^\ContextOne_i)$ and $\mathcal{V}_{1,2}^{1,1}(\mathsf{x}^\ContextTwo_i)$, as illustrated by the dark gray segment of the 45-degree line. Panel (b): Type-$\omega$ decision makers employ $U_{\omega_i}$ for risk assessments in both contexts. They will choose the bundle $\mathcal{I}_{1,1}$ if the parameter $\omega_i$ is below both $\mathcal{W}_{2,1}^{1,1}(\mathsf{x}^\ContextOne_i)$ and $\mathcal{W}_{1,2}^{1,1}(\mathsf{x}^\ContextTwo_i)$, as illustrated by the dark gray segment of the 45-degree line. Panel (c): With context-dependent preference, the decision maker will choose $\mathcal{I}_{1,1}$ if $\nu_i$ is below $\mathcal{V}_{2,1}^{1,1}(\mathsf{x}^\ContextOne_i)$ and $\omega_i$ is below $\mathcal{W}_{1,2}^{1,1}(\mathsf{x}^\ContextTwo_i)$, as illustrated by the dark gray rectangular area. 
\end{flushleft}
\end{figure}

From the above definitions, type-$\nu$ decisions makers will choose bundle $\mathcal{I}_{1,1}$ if and only if $\nu_i\leq \mathcal{V}^{1,1}_{2,1}(\mathsf{x}^\ContextOne_i)$ and $\nu_i\leq \mathcal{V}^{1,1}_{1,2}(\mathsf{x}^\ContextTwo_i)$, and similarly for type-$\omega$ decision makers. In other words, the choice probability satisfies the following:
\begin{align}\label{eq:full consideration original model}
    \mathbb{P}\Big[\mathcal{I}_{1,1}\text{ chosen}\Big|\mathsf{x}^\ContextOne_i,\mathsf{x}^\ContextTwo_i\Big]
    &= \alpha \mathbb{P}\Big[ \nu_i \leq \mathcal{V}^{1,1}_{2,1}(\mathsf{x}^\ContextOne_i) \wedge \mathcal{V}^{1,1}_{1,2}(\mathsf{x}^\ContextTwo_i) \Big|\mathsf{x}^\ContextOne_i,\mathsf{x}^\ContextTwo_i \Big]\nonumber\\
    &\qquad + (1-\alpha) \mathbb{P}\Big[ \omega_i \leq \mathcal{W}^{1,1}_{2,1}(\mathsf{x}^\ContextOne_i) \wedge \mathcal{W}^{1,1}_{1,2}(\mathsf{x}^\ContextTwo_i)\Big|\mathsf{x}^\ContextOne_i,\mathsf{x}^\ContextTwo_i \Big]\nonumber\\
    &= \alpha F\Big(\mathcal{V}^{1,1}_{2,1}(\mathsf{x}^\ContextOne_i) \wedge \mathcal{V}^{1,1}_{1,2}(\mathsf{x}^\ContextTwo_i) \Big) + (1-\alpha) G\Big(\mathcal{W}^{1,1}_{2,1}(\mathsf{x}^\ContextOne_i) \wedge \mathcal{W}^{1,1}_{1,2}(\mathsf{x}^\ContextTwo_i)\Big),
\end{align}
where $a\wedge b = \min\{a,b\}$. See Panel (a) and (b) of Figure \ref{fig:fig1} for an illustration. \citet{Barseghyan-Molinari_2023_JBES} establish point identification of $(\alpha,F,G)$ as follows. Take some $\mathsf{v}$ in the support of $F$, and assume we can find a price combination $(\mathsf{x}^\ContextOne_i,\mathsf{x}^\ContextTwo_i$) such that
\begin{align}\label{eq:full consideration original model inequalities}
    \mathsf{v}=\mathcal{V}^{1,1}_{2,1}(\mathsf{x}^\ContextOne_i) < \mathcal{V}^{1,1}_{1,2}(\mathsf{x}^\ContextTwo_i)\qquad \text{and} \qquad \mathcal{W}^{1,1}_{2,1}(\mathsf{x}^\ContextOne_i) > \mathcal{W}^{1,1}_{1,2}(\mathsf{x}^\ContextTwo_i).
\end{align}
Then, combining \eqref{eq:full consideration original model} and \eqref{eq:full consideration original model inequalities},
\begin{align*}
    \mathbb{P}\Big[\mathcal{I}_{1,1}\text{ chosen}\Big|\mathsf{x}^\ContextOne_i,\mathsf{x}^\ContextTwo_i\Big]
    = \alpha F\Big(\mathcal{V}^{1,1}_{2,1}(\mathsf{x}^\ContextOne_i) \Big) + (1-\alpha) G\Big( \mathcal{W}^{1,1}_{1,2}(\mathsf{x}^\ContextTwo_i)\Big).
\end{align*}
Since an infinitesimal change in $\mathsf{x}^\ContextOne_i$ will not alter the inequalities in \eqref{eq:full consideration original model inequalities}, the following derivative is identifiable at $\mathsf{v}$:
\begin{align*}
    \frac{\partial}{\partial \mathcal{V}^{1,1}_{2,1}(\mathsf{x}^\ContextOne_i)} \mathbb{P}\Big[\mathcal{I}_{1,1}\text{chosen}\Big|\mathsf{x}^\ContextOne_i,\mathsf{x}^\ContextTwo_i\Big]
    &= \alpha f(\mathsf{v}) ,
\end{align*}
where $f$ is the Lebesgue density of $F$. The equality above is intuitive: under \eqref{eq:full consideration original model inequalities}, type-$\nu$ marginal decision makers will be more sensitive to price changes in context $\ContextOne$, while type-$\omega$ marginal decision makers react to prices changes in context $\ContextTwo$. Therefore, a change in $\mathsf{x}^\ContextOne_i$ will affect the threshold $\mathcal{V}^{1,1}_{2,1}(\mathsf{x}^\ContextOne_i)$, which in turn affects the fraction of type-$\nu$ decision makers who will choose the $\mathcal{I}_{1,1}$ bundle. Another key observation is that the threshold function, $\mathsf{x}^\ContextOne_i \mapsto \mathcal{V}^{1,1}_{2,1}(\mathsf{x}^\ContextOne_i)$, can be computed once the analyst has chosen the utility function class $\{ U_\nu:\nu\in[0,\bar{\nu}] \}$; that is, we can directly exploit the variation in the threshold $\mathcal{V}^{1,1}_{2,1}(\mathsf{x}^\ContextOne_i)$.

If the conditions in \eqref{eq:full consideration original model inequalities} are met for all $\mathsf{v}\in [0,\bar{\nu}]$, then $f(\cdot)$ and the mixing probability $\alpha$ are identifiable. An analogous argument can be used to identify $g(\cdot)$ (the density of $G$). This result is formally established in Theorem 3.1 of \cite{Barseghyan-Molinari_2023_JBES}.

%%%%%%%%%%%%%%%%%%%%%%%%%%%%%%%%%%%%%%%%%%%%%%%
%% Context-Dependent Risk Assessment
%%%%%%%%%%%%%%%%%%%%%%%%%%%%%%%%%%%%%%%%%%%%%%%
\subsection{Context-Dependent Risk Assessment}\label{sec: Context-Dependent Mixing Probabilities}

As an attempt to allow the preference to depend on the context (choice problem), assume the population consists of three types:
\begin{itemize}
    \item[(i)] type-$\nu$ individuals always employ the utility function $U_{\nu_i}$ for decision making, where $\nu_i\sim F$;
    \item[(ii)] type-$\omega$ individuals always employ the utility function $U_{\omega_i}$ for decision making, where $\omega_i\sim G$;
    \item[(iii)] individuals of the last type employ context-dependent risk assessment, that is, they use different utility functions, $U_{\nu_i}$ and $U_{\omega_i}$ for decision making in context $\ContextOne$ and $\ContextTwo$, respectively.
\end{itemize}
The unknown proportions of the three types are $\alpha$, $\beta$, and $1-\alpha-\beta$, respectively. This extension can also be understood as context-dependent mixing probabilities, since now the fraction of decision makers employing the utility function $U_{\nu_i}$ will be context specific: $1-\beta$ for context $\ContextOne$ and $\alpha$ for context $\ContextTwo$. 

Since decision makers of the third type are equipped with two (random) utility functions, we have to specify how the random utilities are generated. This is done in the following assumption. 

\begin{Assumption}\label{Assumption: context dependent preference}
For decision makers employing different utility functions for the two contexts, their random utilities are generated from some joint distribution $C(F(\nu),G(\omega))$, where $F$ and $G$ are the marginal distribution of $\nu_i$ and $\omega_i$ and $C(\cdot,\cdot)$ is a continuously differentiable copula function. \qed
\end{Assumption}

The choice behaviors of the first two groups have been analyzed previously. For individuals of the third type that we just introduced, they will pick bundle $\mathcal{I}_{1,1}$ if $\nu_i\leq \mathcal{V}^{1,1}_{2,1}(\mathsf{x}^\ContextOne_i)$ and $\omega_i \leq \mathcal{W}^{1,1}_{1,2}(\mathsf{x}^\ContextTwo_i)$. By Assumption \ref{Assumption: context dependent preference}, 
\begin{align}
    \mathbb{P}\Big[\mathcal{I}_{1,1}\text{ chosen}\Big|\mathsf{x}^\ContextOne_i,\mathsf{x}^\ContextTwo_i\Big]
    &= \alpha F\Big(\mathcal{V}^{1,1}_{2,1}(\mathsf{x}^\ContextOne_i) \wedge \mathcal{V}^{1,1}_{1,2}(\mathsf{x}^\ContextTwo_i) \Big) + \beta G\Big(\mathcal{W}^{1,1}_{2,1}(\mathsf{x}^\ContextOne_i) \wedge \mathcal{W}^{1,1}_{1,2}(\mathsf{x}^\ContextTwo_i)\Big)\nonumber\\
    &\qquad+ (1-\alpha-\beta) C\Big(F\Big(\mathcal{V}^{1,1}_{2,1}(\mathsf{x}^\ContextOne_i)\Big), G\Big(\mathcal{W}^{1,1}_{1,2}(\mathsf{x}^\ContextTwo_i)\Big)\Big).\label{eq: context-Dependent choice probability}
\end{align}
See Panel (c) of Figure \ref{fig:fig1} for an illustration.

In this model of context-dependent risk assessment, we can point identify the mixing probabilities and the distribution of the risk parameters, $(\alpha,\beta,F,G,C)$, following the core idea in \cite{Barseghyan-Molinari_2023_JBES}. Consider some $\mathsf{v}$ in the support of $F$ where one can find price combinations such that the following holds:
\begin{align}\label{eq:context dependent preference, identification inequality}
    \mathsf{v} = \mathcal{V}^{1,1}_{2,1}(\mathsf{x}^\ContextOne_i) = \mathcal{V}^{1,1}_{1,2}(\mathsf{x}^\ContextTwo_i)
    \qquad \text{and} \qquad 
    \mathcal{W}^{1,1}_{2,1}(\mathsf{x}^\ContextOne_i) > \mathcal{W}^{1,1}_{1,2}(\mathsf{x}^\ContextTwo_i).
\end{align}
Since the threshold functions, $\mathcal{V}^{\ell,q}_{k,r}(\cdot)$ and $\mathcal{W}^{\ell,q}_{k,r}(\cdot)$ are continuous, it is possible to break the equality $\mathcal{V}^{1,1}_{2,1}(\mathsf{x}^\ContextOne_i) = \mathcal{V}^{1,1}_{1,2}(\mathsf{x}^\ContextTwo_i)$ by slight variation in $\mathsf{x}^\ContextTwo_i$ without affecting the second constraint in \eqref{eq:context dependent preference, identification inequality}. That is,
\begin{alignat*}{2}
    &\lim_{\mathcal{V}^{1,1}_{1,2}(\mathsf{x}^\ContextTwo_i)\downarrow \mathsf{v}}\ \frac{\partial}{\partial \mathcal{V}^{1,1}_{2,1}(\mathsf{x}^\ContextOne_i)} \mathbb{P}\Big[\mathcal{I}_{1,1}\text{ chosen}\Big|\mathsf{x}^\ContextOne_i,\mathsf{x}^\ContextTwo_i\Big]
    = \Bigg\{\alpha + &&(1-\alpha-\beta)C_1\Bigg(F(\mathsf{v}), G\Big(\mathcal{W}^{1,1}_{1,2}(\mathsf{x}^\ContextTwo_i)\Big)\Bigg) \Bigg\}  f(\mathsf{v}) ,\\
    &\lim_{\mathcal{V}^{1,1}_{1,2}(\mathsf{x}^\ContextTwo_i)\uparrow \mathsf{v}}\ \frac{\partial}{\partial \mathcal{V}^{1,1}_{2,1}(\mathsf{x}^\ContextOne_i)} \mathbb{P}\Big[\mathcal{I}_{1,1}\text{ chosen}\Big|\mathsf{x}^\ContextOne_i,\mathsf{x}^\ContextTwo_i\Big]
    = \Bigg\{&&(1-\alpha-\beta)C_1\Bigg(F(\mathsf{v}), G\Big(\mathcal{W}^{1,1}_{1,2}(\mathsf{x}^\ContextTwo_i)\Big)\Bigg) \Bigg\}f(\mathsf{v}),
\end{alignat*}
where $C_1(\cdot,\cdot)$ is the derivative of the copula function with respect to its first argument. Therefore, the discontinuity-in-derivative formula gives
\begin{align*}
    \lim_{\mathcal{V}^{1,1}_{1,2}(\mathsf{x}^\ContextTwo_i)\downarrow \mathsf{v}}\ \frac{\partial}{\partial \mathcal{V}^{1,1}_{2,1}(\mathsf{x}^\ContextOne_i)} \mathbb{P}\Big[\mathcal{I}_{1,1}\text{ chosen}\Big|\mathsf{x}^\ContextOne_i,\mathsf{x}^\ContextTwo_i\Big] - \lim_{\mathcal{V}^{1,1}_{1,2}(\mathsf{x}^\ContextTwo_i)\uparrow \mathsf{v}}\ \frac{\partial}{\partial \mathcal{V}^{1,1}_{2,1}(\mathsf{x}^\ContextOne_i)} \mathbb{P}\Big[\mathcal{I}_{1,1}\text{ chosen}\Big|\mathsf{x}^\ContextOne_i,\mathsf{x}^\ContextTwo_i\Big] = \alpha f(\mathsf{v}),
\end{align*}
which provides identification of $f(\cdot)$ and the mixing probability $\alpha$ if there is enough variation in prices such that \eqref{eq:context dependent preference, identification inequality} can be constructed for all $\mathsf{v}$ in the support. We summarize our findings in the following theorem. 
\begin{thm}
    Let Assumptions 2.1, 2.3, and 2.4 in \cite{Barseghyan-Molinari_2023_JBES} and our Assumption \ref{Assumption: context dependent preference} hold. In addition, assume there is enough variation in prices, $\mathsf{x}^\ContextOne_i$ and $\mathsf{x}^\ContextTwo_i$, such that \eqref{eq:context dependent preference, identification inequality} is feasible for all $\mathsf{v}$ in the support of $F$. Then $\alpha$ and $F$ are identified. \qed
\end{thm}

By symmetry, the same argument applied to $\mathcal{W}^{1,1}_{2,1}(\cdot)$ and $\mathcal{W}^{1,1}_{1,2}(\cdot)$ can be used to point identify $(\beta,G)$, and subsequently the copula function $C$.

%%%%%%%%%%%%%%%%%%%%%%%%%%%%%%%%%%%%%%%%%%%%%%%
%%%%%%%%%%%%%%%%%%%%%%%%%%%%%%%%%%%%%%%%%%%%%%%
%% Context-Dependent Risk Assessment with Limited Consideration
%%%%%%%%%%%%%%%%%%%%%%%%%%%%%%%%%%%%%%%%%%%%%%%
%%%%%%%%%%%%%%%%%%%%%%%%%%%%%%%%%%%%%%%%%%%%%%%
\section{Context-Dependent Risk Assessment with Limited Consideration}\label{sec: Context-Dependent Mixing Probabilities under Limited Consideration}

To allow for context-dependent risk assessments with limited consideration is a nontrivial task. In particular, the notation quickly becomes cumbersome. In this section, we thus make a simplifying assumption on the support of the consideration sets.

\begin{Assumption}\label{Assu: Random consideration, support condition}
   Let $\RandomConsideration(\cdot)$ be the probability measure representing random consideration. Let
   \begin{alignat*}{2}
       &\RandomConsideration_{\{1,2\}\times \{1,2\}} &&:=\ \RandomConsideration(\{\mathcal{I}_{\ell,q}:\ \ell,q = 1,2\}),\\
       &\RandomConsideration_{\{\ell\}\times \{1,2\}} &&:=\ \RandomConsideration(\{\mathcal{I}_{\ell,q}:\ q = 1,2\}),\quad \ell=1,2,\\
       &\RandomConsideration_{\{1,2\}\times \{q\}} &&:=\ \RandomConsideration(\{\mathcal{I}_{\ell,q}:\ \ell = 1,2\}),\quad q=1,2,\\
       &\RandomConsideration_{\{\ell\}\times \{q\}} &&:=\ \RandomConsideration(\{\mathcal{I}_{\ell,q}\}),\quad \ell,q = 1,2.
   \end{alignat*}
   Then,
   \[
    \RandomConsideration_{\{1,2\}\times \{1,2\}} + \sum_{\ell=1,2} \RandomConsideration_{\{\ell\}\times \{1,2\}} + \sum_{q=1,2} \RandomConsideration_{\{1,2\}\times \{q\}} + \underset{\ell,q = 1,2}{\sum\sum} \RandomConsideration_{\{\ell\}\times \{q\}} = 1.
   \]
   \qed
\end{Assumption}

For example, $\RandomConsideration_{\{1,2\}\times \{1,2\}}$ is the probability of full attention (i.e., the chance that the decision maker pays attention to both options, 1 and 2, in both contexts). Similarly, $\RandomConsideration_{\{1,2\}\times \{2\}}$ is the probability that she pays attention to both 1 and 2 in $\ContextOne$ but only 2 in $\ContextTwo$. The assumption rules out consideration sets such as $\{ \mathcal{I}_{1,1}, \mathcal{I}_{2,2} \}$, which simplifies the notation and presentation. Assumption \ref{Assu: Random consideration, support condition} is not necessary for our identification results. In particular, if consideration bundles such as $\{ \mathcal{I}_{1,1}, \mathcal{I}_{2,2} \}$ were allowed, then we would only need to specify how individuals of the third type (see Section \ref{sec: Context-Dependent Mixing Probabilities}) make decisions. 

Under Assumption \ref{Assu: Random consideration, support condition}, the choice probability for bundle $\mathcal{I}_{1,1}$ can be written as
\begin{alignat*}{3}
     \nonumber\mathbb{P}\Big[\mathcal{I}_{1,1}\text{ chosen}\Big|\mathsf{x}^\ContextOne_i,\mathsf{x}^\ContextTwo_i\Big] =\alpha &\bigg\{\ \    
    &&\RandomConsideration_{\{1,2\}\times \{1,2\}} F\Big(\mathcal{V}^{1,1}_{2,1}(\mathsf{x}^\ContextOne_i) \wedge \mathcal{V}^{1,1}_{1,2}(\mathsf{x}^\ContextTwo_i) \Big) &&\ \\
    &\ + &&\RandomConsideration_{\{1,2\}\times \{1\}} F\Big(\mathcal{V}^{1,1}_{2,1}(\mathsf{x}^\ContextOne_i) \Big)
    \ +\ \RandomConsideration_{\{1\}\times \{1,2\}} F\Big(\mathcal{V}^{1,1}_{1,2}(\mathsf{x}^\ContextTwo_i) \Big)
    \ +\ \RandomConsideration_{\{1\}\times \{1\}} \ \    &&\Bigg\}\\
    +\beta &\bigg\{\   
    &&\RandomConsideration_{\{1,2\}\times \{1,2\}} G\Big(\mathcal{W}^{1,1}_{2,1}(\mathsf{x}^\ContextOne_i) \wedge \mathcal{W}^{1,1}_{1,2}(\mathsf{x}^\ContextTwo_i) \Big) &&\ \\
        &\ + &&\RandomConsideration_{\{1,2\}\times \{1\}} G\Big(\mathcal{W}^{1,1}_{2,1}(\mathsf{x}^\ContextOne_i) \Big)
        \ +\ \RandomConsideration_{\{1\}\times \{1,2\}} G\Big(\mathcal{W}^{1,1}_{1,2}(\mathsf{x}^\ContextTwo_i) \Big)
        \ +\ \RandomConsideration_{\{1\}\times \{1\}} \ \ &&\Bigg\}\\
    +(1-\alpha-\beta) &\bigg\{\   
    &&\RandomConsideration_{\{1,2\}\times \{1,2\}} C\Bigg(F\Big(\mathcal{V}^{1,1}_{2,1}(\mathsf{x}^\ContextOne_i) \Big),G\Big(    \mathcal{W}^{1,1}_{1,2}(\mathsf{x}^\ContextTwo_i) \Big)\Bigg)&\ \\
        &\ + &&\RandomConsideration_{\{1,2\}\times \{1\}} F\Big(\mathcal{V}^{1,1}_{2,1}(\mathsf{x}^\ContextOne_i) \Big)
        \ +\ \RandomConsideration_{\{1\}\times \{1,2\}} G\Big(\mathcal{W}^{1,1}_{1,2}(\mathsf{x}^\ContextTwo_i) \Big)
        \ +\ \RandomConsideration_{\{1\}\times \{1\}} \ \ &&\Bigg\}.
\end{alignat*}
As in \cite{Barseghyan-Molinari_2023_JBES}, it is also possible to allow the random consideration, $\RandomConsideration(\cdot)$, to depend on the type of decision makers. In other words, the three types of decision makers (Section \ref{sec: Context-Dependent Mixing Probabilities}) will be equipped with different random consideration measures, say $\RandomConsideration^{\text{(i)}}(\cdot)$, $\RandomConsideration^{\text{(ii)}}(\cdot)$, and $\RandomConsideration^{\text{(iii)}}(\cdot)$. We abstract away from this generalization to save notation. 

Now assume \eqref{eq:context dependent preference, identification inequality} is possible for some $\mathsf{v}$ in the support of $F$, 
then the discontinuity-in-derivative formula yields:
\begin{align*}
\lim_{\mathcal{V}^{1,1}_{1,2}(\mathsf{x}^\ContextTwo_i)\downarrow \mathsf{v}}\ \frac{\partial}{\partial \mathcal{V}^{1,1}_{2,1}(\mathsf{x}^\ContextOne_i)} \mathbb{P}\Big[\mathcal{I}_{1,1}\text{ chosen}\Big|\mathsf{x}^\ContextOne_i,\mathsf{x}^\ContextTwo_i\Big] - \lim_{\mathcal{V}^{1,1}_{1,2}(\mathsf{x}^\ContextTwo_i)\uparrow \mathsf{v}}\ \frac{\partial}{\partial \mathcal{V}^{1,1}_{2,1}(\mathsf{x}^\ContextOne_i)} \mathbb{P}\Big[\mathcal{I}_{1,1}\text{ chosen}\Big|\mathsf{x}^\ContextOne_i,\mathsf{x}^\ContextTwo_i\Big] &= \alpha\RandomConsideration_{\{1,2\}\times \{1,2\}}f(\mathsf{v}).
\end{align*}
We summarize the identification result in the following theorem. 

\begin{thm}
    Let Assumptions 2.1, 2.3, and 2.4 in \cite{Barseghyan-Molinari_2023_JBES} and our Assumptions \ref{Assumption: context dependent preference} and \ref{Assu: Random consideration, support condition} hold. In addition, assume there is enough variation in prices, $\mathsf{x}^\ContextOne_i$ and $\mathsf{x}^\ContextTwo_i$, such that \eqref{eq:context dependent preference, identification inequality} is feasible for all $\mathsf{v}$ in the support of $F$. Then $\alpha\RandomConsideration_{\{1,2\}\times \{1,2\}}$ and $F$ are identified. \qed
\end{thm}

%%%%%%%%%%%%%%%%%%%%%%%%%%%%%%%%%%%%%%%%%%%%%%%
%%%%%%%%%%%%%%%%%%%%%%%%%%%%%%%%%%%%%%%%%%%%%%%
%% Conclusion
%%%%%%%%%%%%%%%%%%%%%%%%%%%%%%%%%%%%%%%%%%%%%%%
%%%%%%%%%%%%%%%%%%%%%%%%%%%%%%%%%%%%%%%%%%%%%%%
\section{Conclusion}\label{sec: Conclusion}

\citet{Barseghyan-Molinari_2023_JBES} introduced and studied an interesting model of decision-making under risk, allowing for unobserved heterogeneity in utility functions and consideration set formation, where the mixing probability parameter determining the risk profile for each decision maker is unknown but context-independent. They provided insightful identification results of parameters of interest (the distribution of the random coefficient distributions and the context-independent mixing probability). Motivated by the abundant theoretical and empirical behavioral literature, we enhanced \citet{Barseghyan-Molinari_2023_JBES}'s model to allow for context-dependent random utility. More precisely, we allowed for the mixing probability entering the finite mixture of preference types to be context-dependent. We then built on their identification approach to establish semi-nonparametric point identification of parameters of interest. \bigskip

\section*{Acknowledgments} We thank Francesca Molinari and the participants at the 2023 ASSA meetings (JBES Session: Risk Preference Types, Limited Consideration, and Welfare) for comments.

\section*{Funding} Cattaneo gratefully acknowledges financial support from the National Science Foundation through grants SES-1947805 and SES-2241575.

\section*{Conflict of Interest} The authors report there are no competing interests to declare.

\bibliographystyle{econometrica}
\bibliography{CMM_2023_JBES--comment-bib}

\end{document}